\documentclass[prl,twocolumn,aps]{revtex4-1}

\usepackage{epsfig}
\usepackage{graphicx}
\usepackage{color}

\begin{document}

\title{Vortex arrays in neutral trapped Fermi gases throughout the BCS-BEC crossover}

\author{S. Simonucci,$^{1,2}$  P. Pieri,$^{1}$  and G. Calvanese Strinati$^{1,2}$}

\affiliation{$^{1}$Division of Physics, School of Science and Technology, Universit\`{a} di Camerino, 62032 Camerino (MC), Italy  \\
                $^{2}$INFN, Sezione di Perugia, 06123 Perugia (PG), Italy}



\begin{abstract}
{\bf Vortex arrays in type-II superconductors admit the translational symmetry of an infinite system.
There are cases, however, like ultra-cold trapped Fermi gases and the crust of neutron stars, where finite-size effects make it quite more complex to account for the geometrical arrangement of vortices. 
Here, we self-consistently generate these arrays of vortices at zero and finite temperature through a microscopic description of the non-homogeneous superfluid based on a differential equation for the local order parameter, obtained by coarse graining the Bogoliubov-de Gennes (BdG) equations.
In this way, the strength of the inter-particle interaction is varied along the BCS-BEC crossover, from largely overlapping Cooper pairs in the BCS limit to dilute composite bosons in the BEC limit.
Detailed comparison with two landmark experiments on ultra-cold Fermi gases, aimed at revealing the presence of the superfluid phase, brings out several features that makes them relevant for other systems in nature as well.}
\end{abstract} 

\maketitle

The Meissner effect provides evidence of the superconducting phase below the critical temperature $T_{c}$ \cite{Tinkham-1975,Abrikosov-1988}.
In particular, in type-II superconductors the flux of an applied magnetic field $H$ is partially or totally expelled from the material, depending on its strength.
Two critical values of $H$ are distinguished, such that for $0 \le H < H_{c_{1}}$ the magnetic flux is totally expelled while for $H_{c_{1}} \le H < H_{c_{2}}$ the system allows for the penetration of an ordered array of vortices.
Both $ H_{c_{1}}$ and $H_{c_{2}}$ depend on the temperature $T$, making $T_{c}$ depend on $H$.

These considerations apply to large systems containing an (essentially) infinite number of particles, such that also the number of vortices can be considered infinite for all practical purposes.
There exist, however, systems with a \emph{finite number of particles} $N$ where superfluid behavior occurs.
They include nuclei \cite{Schuck-2004} with $N \! \sim \!10^{2}$ and ultra-cold trapped Fermi gases \cite{Ketterle-Zwierlein-2007} typically with $N \! \sim \! 10^{4}\!-\!10^{6}$.
In these systems, orbital effects, which in superconducting materials are associated with an applied magnetic field, result by rotating the sample as a whole \cite{Fetter-2009}
or through synthetic gauge fields \cite{Lin-2009a,Lin-2009b,Dalibard-2011}.

In particular, ultra-cold Fermi gases have attracted special interest recently \cite{Bloch-2008,Giorgini-2008}, because their inter-particle interaction can be varied almost at will 
via the use of Fano-Feshbach  resonances \cite{Fano-1,Fano-2,Feshbach,Grimm-2010}.
These circumstances make the Meissner-like effect in these systems depend also on the coupling parameter $(k_{F} a_{F})^{-1}$ (see Methods).
In addition, surface effects may become prominent in systems with finite $N$, since a finite fraction of the sample can become normal in the ``outer'' region due to rotation while the ``inner'' region remains superfluid.
The Meissner-like effect in these bounded systems can be connected with related effects occurring in nuclei where the moment of inertia gets quenched by superfluidity \cite{Schuck-2004}, and in the inner crust of neutron stars where array of vortices form close to the surface \cite{Broglia,Haensel}. 

Two experiments were performed by setting an ultra-cold trapped Fermi gas into rotation, aiming at revealing the presence of the superfluid phase.
In the first experiment \cite{Ketterle-2005}, arrays of vortices were generated at low temperature throughout the BCS-BEC crossover, providing the first direct evidence for the occurrence of the superfluid phase in these systems. 
In the second experiment \cite{Grimm-2011}, the superfluid behavior was revealed by the quenching of the moment of inertia, which was measured at unitarity (where $(k_{F} a_{F})^{-1}=0$) for increasing temperature until its classical value of the normal state was reached.
In both experiments, the two lowest hyperfine atomic levels were equally populated for a gas of ($^{6}$Li) Fermi atoms contained in an ellipsoidal trap of radial $\nu_{r}$ and axial $\nu_{z}$ frequencies (see Methods), and the system was assumed to reach a (quasi) equilibrium situation at any given rotation frequency $\nu$.
These experiments complement each other, in that they involve determining the total angular momentum and controlling the appearance of vortices.

Despite the role played by these experiments, no theoretical account exists for the structure of complex arrays of quantum vortices
in a rotating superfluid with a finite number of particles in realistic traps as functions of inter-particle coupling and temperature, nor for the related behavior of the moment of inertia.
This is because accounting theoretically for these complex arrays of vortices across the BCS-BEC crossover becomes computationally prohibitive when relying on a standard tool like the BdG equations, which extend the BCS mean field to nonuniform situations \cite{BdG}.
Here, we tackle this difficult computational problem by resorting to a novel differential equation for the local gap parameter $\Delta(\mathbf{r})$ at position $\mathbf{r}$, through which it is possible to find solutions in a rotating trap also for large numbers of vortices.
This equation was introduced in ref.\cite{SS-LPDA} (see Methods) and rests on the assumption that the phase $\varphi(\mathbf{r})$ of the complex function $\Delta(\mathbf{r}) = |\Delta(\mathbf{r})| \, e^{i \varphi(\mathbf{r})}$ varies more rapidly than the magnitude $|\Delta(\mathbf{r})|$.
Accordingly, this equation was referred to as a Local Phase Density Approximation (LPDA) to the BdG equations.
In the following, we shall determine the behavior of the gap parameter, density, and current for a rotating trapped Fermi gas using the LPDA equation, while varying the thermodynamic variables $(k_{F} a_{F})^{-1}$ and $T$, as well as the angular frequency $\Omega = 2 \pi \nu$.

Our theoretical analysis highlights several features which are of interest also to other branches of physics, such as: 
(i) The deviations from Feynman's theorem for the vortex spacing in an array;
(ii) The bending of a vortex filament at the surface of the cloud; 
(iii) The shape of the central vortex surrounded by a finite number of vortices;
(iv) The emergence of a bi-modal distribution for the density profile in the superfluid phase;
(v) The orbital analog of a breached-pair phase;
(vi) The yrast effect similar to that occurring in nuclei;
(vii) The temperature vs frequency phase diagram in a finite-size system. 

Previous work on vortices in Fermi gases considered a single vortex with cylindrical symmetry at zero temperature, either within a BdG approach in the BCS regime \cite{Feder-2003} and
across the BCS-BEC crossover \cite{Levin-2006,Randeria-2006}, or including energy-density-functional corrections at unitarity \cite{Bulgac-2003}.
Vortex lattices were considered only for a strictly two-dimensional geometry, at zero temperature in the BCS regime within a static \cite{Feder-2004} and a dynamic \cite{Castin-2006} BdG approach, and close to $T_{c}$ at unitarity within a Ginzburg-Landau approach \cite{Gao-2008}.
None of these works could address a comparison with the experimental data of refs.\cite{Ketterle-2005,Grimm-2011}, nor bring out the physical features highlighted above.

We first consider the conditions of the experiment of ref.\cite{Ketterle-2005} with $N= 2 \times 10^{6}$ and an aspect ratio $2.5$ (see Methods), with an atomic cloud large enough to support many vortices at low temperature.
In this experiment, the trap was set in rotation at an optimal  stirring frequency $\nu_{\mathrm{opt}} = 45$Hz.

\begin{figure}[ht!]
\includegraphics[angle=0,width=7.0cm]{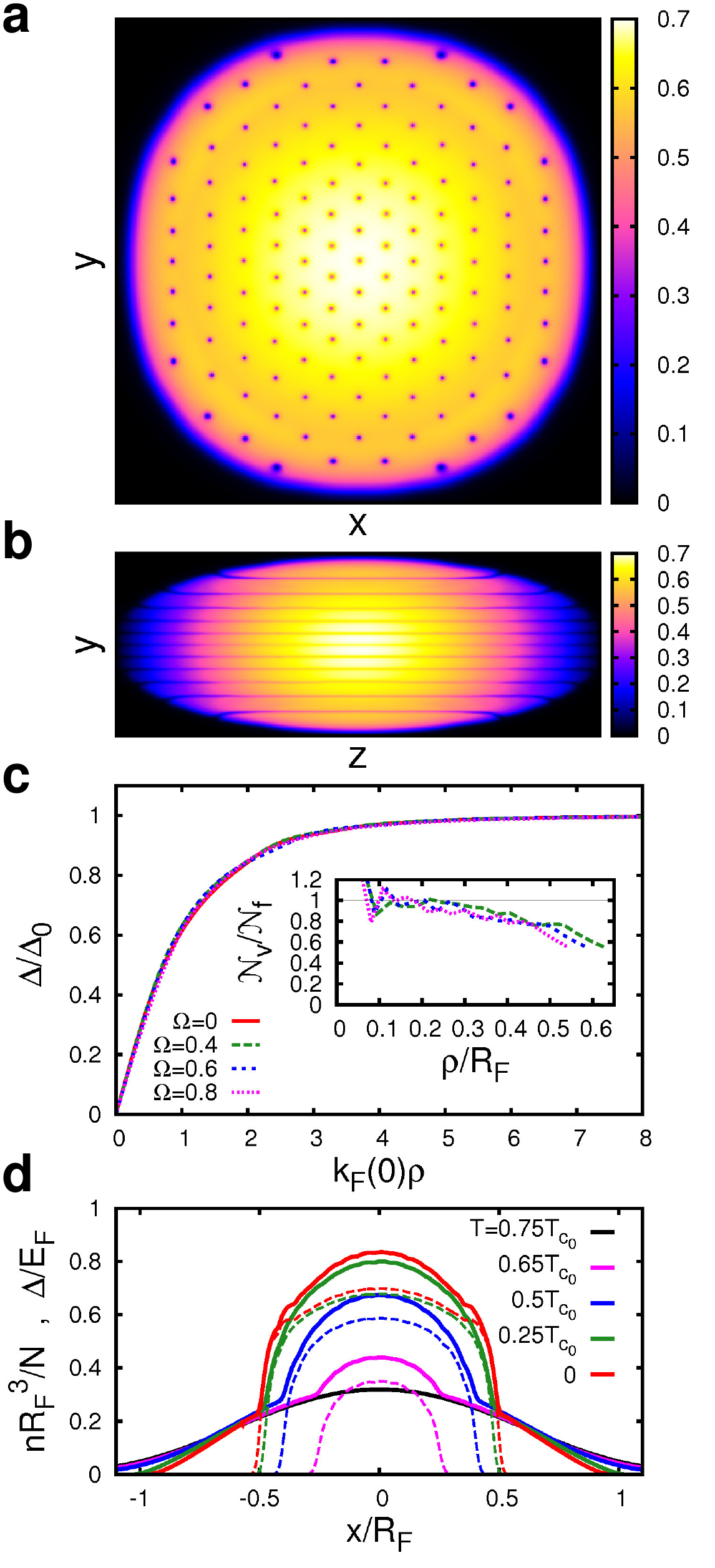}
\caption{\begin{small}{\bf Arrays of vortices at unitarity for $\mathbf{\Omega = 0.8\, \Omega_{r}}$.}  
               Magnitude of the gap parameter obtained from the self-consistent solution of the LPDA equation at $T=0$, seen from the top ({\bf a}) and from the side ({\bf b}). 
              {\bf c}. Radial profile of the (magnitude of the) gap parameter (in units of its value $\Delta_{0}$ in between two adjacent vortices) inside the central vortex in the trap 
              for various angular frequencies (in units of $\Omega_{r}$) at $T=0$. 
              The inset compares the vortex count over an increasing superfluid section of the cloud with the prediction of Feyman's theorem
              (where $R_{F} = [2 E_{F} / (m \Omega_{0}^{2})]^{1/2}$ is the Thomas-Fermi radius, with the Fermi energy $E_{F}$ and 
              $\Omega_{0}$ given in Methods).
              {\bf d}. Profile of the (magnitude of the) gap parameter (broken lines) and of the density (full lines) for different temperatures  (in units of  the critical temperature $T_{c_{0}}$ for $\Omega=0$). \end{small}}              
\label{Figure-1}
\end{figure}

Figure \ref{Figure-1} shows false colour images of the array of vortices obtained theoretically at unitarity and zero temperature for the experimental set up of ref.\cite{Ketterle-2005} with  $\Omega = \Omega_{\mathrm{opt}} = 0.8\, \Omega_{r}$, as seen from the top at $z=0$ [panel (a)] and from the side at $x=0$ [panel (b)] (where $ \Omega_{r}=2\pi\nu_r$).
In Fig. \ref{Figure-1}a, the array contains $137$ vortices, whose locations are not fixed \emph{a priori} but are self-consistently determined by solving the LPDA equation (see Supplementary Information).
The spacing among vortices in the triangular array increases away from the trap center.
Near the trap center this spacing is found to be $[\hbar \pi / (\sqrt{3} m \Omega)]^{1/2}$, consistent with the Feynman's theorem for the number of vortices per unit area $n_{\mathrm{v}} = 2 m \Omega / \pi \hbar$ which applies to an infinite array of vortices \cite{NP-1990} (where $\hbar$ is the Planck constant $h$ divided by $2 \pi$).
In Fig. \ref{Figure-1}b, 11 vortex filaments are distinguished and appear to bend away from the $z$ (axial) direction upon approaching the boundary of the cloud, so as to meet the surface perpendicularly with no current leaking out of the cloud
(this feature was noted previously for a single vortex in an elongated trapped bosonic cloud \cite{Castin-2003}). 
Figure \ref{Figure-1}c shows the radial profile of the (magnitude of the) gap parameter inside the central vortex for several values of the angular frequency $\Omega$, whose shape appears to be unmodified once  normalized to the ``asymptotic'' value $\Delta_{0}$ in between two adjacent vortices and the radial distance $\rho = \sqrt{x^{2}+y^{2}}$ is expressed in terms of the local Fermi wave vector $k_{F}(0) = [3 \pi^{2} n(0)]^{1/3}$ where $n(0)$ is the density at the trap center. 
[Specifically, $\Delta_{0} = (0.87,0.83,0.77,0.68) \, E_{F}$ for $\Omega = (0.0,0.4,0.6,0.8) \, \Omega_{r}$.]
This is because the healing length of the central vortex (which is about $2 k_{F}(0)^{-1}$ in Fig. \ref{Figure-1}c) is much smaller than the distance between two adjacent vortices (which is about $24 k_{F}(0)^{-1}$ in Fig. \ref{Figure-1}a where the vortex density is highest).
The inset of Fig. \ref{Figure-1}c shows the $\rho$-dependence of the ratio between the number of vortices $\mathcal{N}_{\mathrm{v}}(\rho)$ obtained numerically within a circle (at $z = 0$) with center at $x=y=0$ and radius $\rho$, and the corresponding number of vortices $\mathcal{N}_{f}(\rho) = n_{\mathbf{v}} \pi \rho^{2}$ expected from Feynman's theorem, for the frequencies  $\Omega = (0.4,0.6,0.8) \, \Omega_{r}$.
Each plot terminates at the boundary $R_{s}$ of the superfluid portion of the cloud (cf. Fig. \ref{Figure-1}d), which is always smaller than $R_{F}$.
One concludes that Feynman's theorem is satisfied, in practice, up to about $R_{s}/2$.
Finally, Fig. \ref{Figure-1}d shows the profiles of the (magnitude of the) gap parameter and density at $z=0$ along  $x$  for $\Omega = 0.8 \, \Omega_{r}$ and various temperatures in the superfluid phase.
A non-negligible portion of the outer normal component of the cloud past $R_{s}$ becomes normal due to rotation, with the result that \emph{a bi-modal distribution} for the density emerges below a certain temperature.
This feature bears analogies with what occurs for a bosonic system, for which the superfluid component emerges near the trap center as a narrow peak out of a broad thermal distribution \cite{DGPS-1999}.
Our results suggest that the emergence of a superfluid component below $T_{c}$ could be detected also for a rotating fermion system directly from the \emph{in situ} density profiles, avoiding the need to expand the cloud when looking for vortex arrays.

\begin{figure}[t]
\includegraphics[angle=0,width=8.0cm]{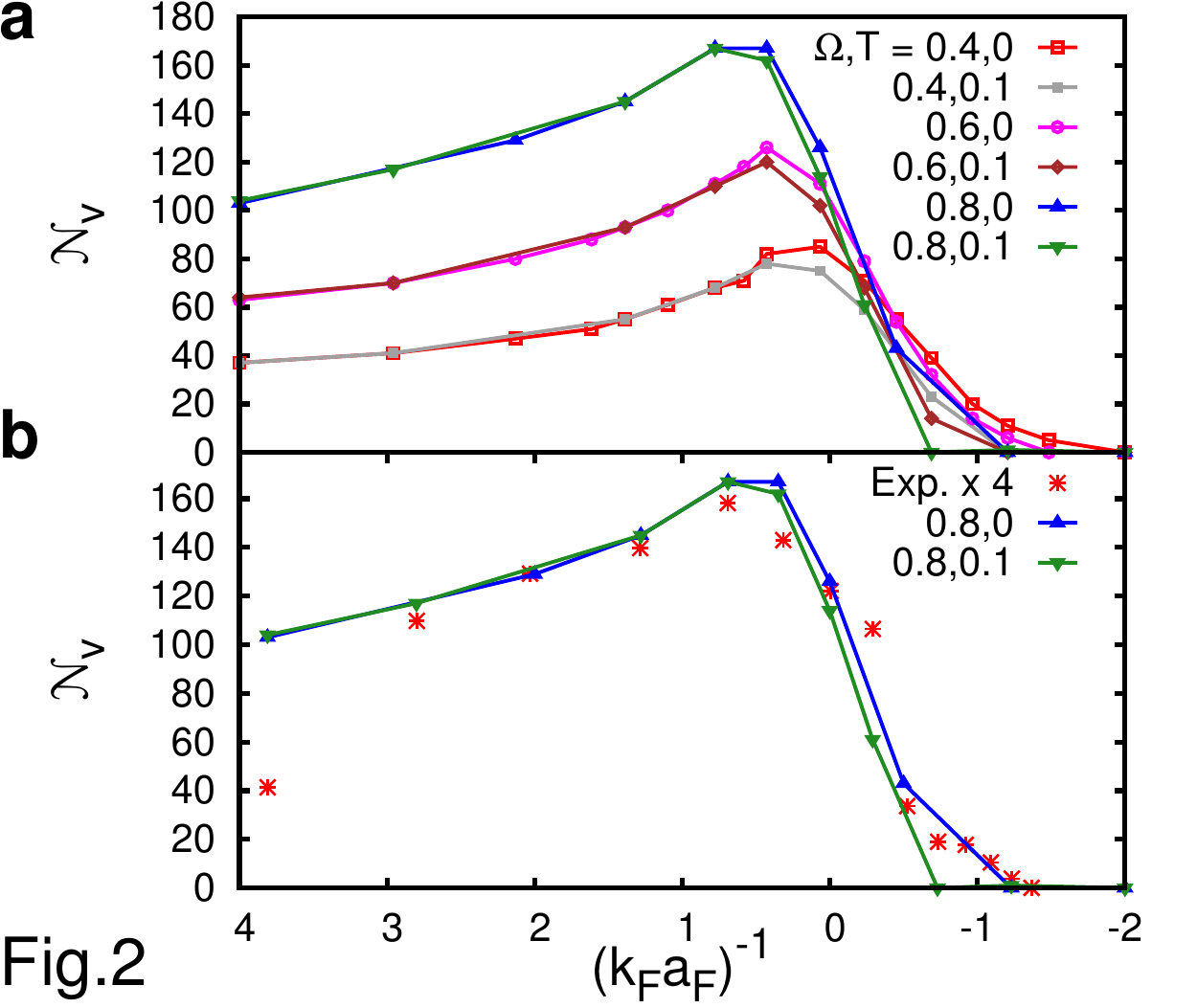}
\caption{\begin{small}{\bf Comparing the number of vortices with the experimental data. a.} Number of vortices $\mathcal{N}_{\mathrm{v}}$ obtained from the solution 
               of the LPDA equation over an extended coupling range across unitarity, for different values of the angular frequency $\Omega$ (in units of 
               $\Omega_{r}$) and of the temperature (in units of the Fermi temperature $T_{F}=E_{F}/k_{B}$, $k_{B}$ being the Boltzmann constant). 
               {\bf b.} The experimental values for $\mathcal{N}_{\mathrm{v}}$ from ref.\cite{Ketterle-2005} multiplied by a factor of four (stars) are compared with the results of the 
               present calculation for $\Omega = 0.8 \, \Omega_{r}$ and two different temperatures (in units of $T_{F}$). 
               \end{small}}
\label{Figure-2}
\end{figure}

When approaching the BCS limit or for increasing $T$, the outer normal component increases in size at the expense of the superfluid component in the inner part of the cloud.
In particular, at $T=0$ the mechanism for activating the outer normal component stems from the presence of the ``classical'' velocity field  
$\mathbf{v}_{\mathrm{n}}(\mathbf{r}) = \mathbf{\Omega} \times \mathbf{r}$ in the argument of the Fermi functions entering the coefficients of the LPDA equation (see Methods).
This outer normal component can accordingly be referred to as an \emph{orbital breached pair} (OBP) phase, in analogy with the breached pair phase for imbalanced spin populations \cite{Wilczek-2003}.
At a given temperature, the spatial extent of the OBP phase increases upon approaching the BCS limit as the magnitude $|\Delta(\mathbf{r})|$ of the gap parameter becomes progressively smaller, in such a way that the total number of vortices drops considerably when approaching this limit.
[A related study of pair-breaking effects in a rotating Fermi gas along the BCS-BEC crossover was made in ref.\cite{Schuck-2008}, albeit in the absence of vortices.]

We have performed our calculations over an extended range of $(k_{F} a_{F})^{-1}$ across unitarity for several values of the angular frequency $\Omega$, and obtained the number of vortices $\mathcal{N}_{\mathrm{v}}$ shown in Fig. \ref{Figure-2}a at $T=0$ and $T=0.1T_{F}$.
We have obtained $\mathcal{N}_{\mathrm{v}}$ from the total circulation $\oint d{\bf \ell} \cdot \nabla \varphi = 2 \pi \mathcal{N}_{\mathrm{v}}$, where the line integral is over a circle which encompasses the whole superfluid portion of the cloud at $z=0$.
For increasing $\Omega$, the maximum of $\mathcal{N}_{\mathrm{v}}$ is found to shift toward the BEC side of unitarity. 
Past this maximum, the temperature has only a minor effect on $\mathcal{N}_{\mathrm{v}}$.
In particular at zero temperature, we find the maximum to occur at $(k_{F} a_{F})^{-1}=(0.13,0.27,0.47)$ for $\Omega=(0.4,0.6,0.8) \, \Omega_{r}$.
In all cases, a rapid decrease of $\mathcal{N}_{\mathrm{v}}$ occurs when approaching the BCS side of unitarity, in accordance with the above argument for the presence of an OBP phase in the outer portion of the cloud.
Figure \ref{Figure-2}b compares our results with the experimental values of $\mathcal{N}_{\mathrm{v}}$ from ref.\cite{Ketterle-2005} at the optimal stirring frequency $\Omega=0.8 \, \Omega_{r}$.
A remarkable agreement is found, provided one multiplies all the experimental data by the \emph{same} factor of four, irrespective of coupling.
[We attribute the discrepancy on the number of vortices at coupling 3.8 to the experimental problems that arise deep in the BEC regime with the relaxation of the highest vibrational state of the molecular potential involved in the Fano-Feshbach resonance of $^{6}\mathrm{Li}$  \cite{Ketterle-2005}.]
The need for this rescaling by a factor of four can be related to our finding (cf. the inset of Fig. \ref{Figure-1}c) that Feynman's theorem is satisfied only in about one-fourth of the area of the cloud, leading us to speculate that only those vortices residing in this central portion of the cloud could experimentally be detected \emph{after} the cloud was expanded. 
This problem might be overcome by taking \emph{in situ} images of the two-dimensional vortex distribution, as was recently done for a Bose gas \cite{Anderson-2014}.
Note in this context also the excellent agreement with the experimental data for the position of the maximum, as it results from Fig. \ref{Figure-2}a for the angular frequency 
$\Omega = 0.8 \, \Omega_{r}$ of the experiment.

We now go on to consider the experiment of ref.\cite{Grimm-2011}, which measured the moment of inertia $\Theta$ of the atomic cloud at unitarity as a function of temperature and showed its progressive quenching as the temperature was lowered below $T_{c}$.
The quenching of $\Theta$ (with respect to its classical value $\Theta_{\mathrm{cl}}$ $-$ see Methods) entails the absence of quantum vortices in the cloud, since these would act to bring the moment of inertia back to $\Theta_{\mathrm{cl}}$ according to Feynman's theorem.
To get an estimate for the value of the frequency $\Omega_{c_{1}}$ at which the first vortex enters the cloud of radial size $R_{s}$, we use the expression 
\vspace{-0.50cm}
 
\begin{equation}
\hbar \Omega_{c_{1}}/E_{F} = 2 (k_{F} R_{s})^{-2} \ln(R_{s}/\xi) \, ,
\label{approximate-Omega-c1}
\end{equation}

\vspace{-0.10cm} 
\noindent
where $\xi$ is the healing length of the vortex. 
This expression is obtained by setting $E - L \Omega = 0$, where $E$ is the energy of an isolated vortex in the laboratory frame and $L$ its total angular momentum \cite{NP-1990}.
The dominant contribution to $E$ originates from the angular kinetic energy outside the vortex core of extent $\xi$.
\begin{figure}[t]
\includegraphics[angle=0,width=8.0cm]{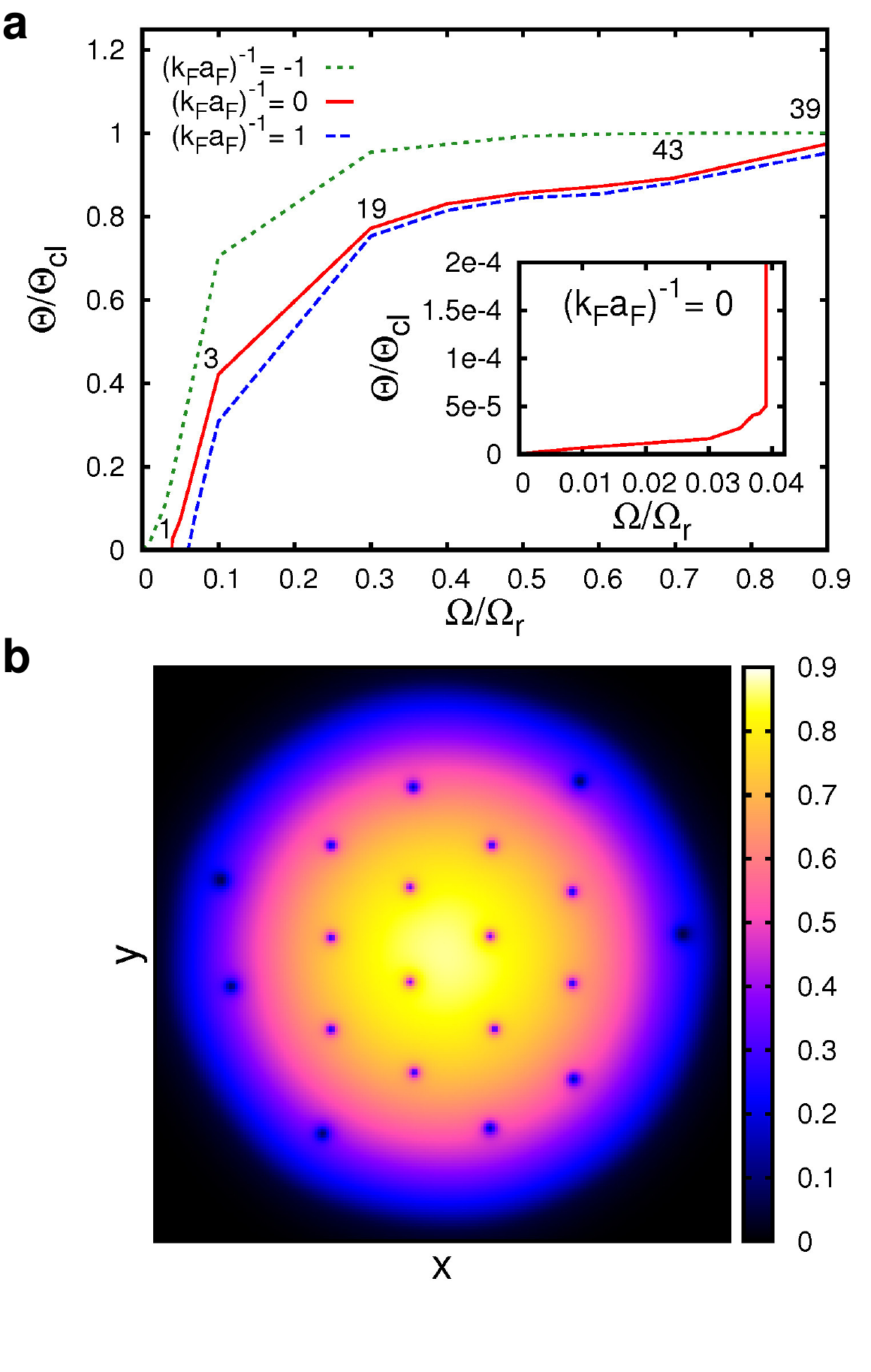}
\caption{\begin{small}{\bf Dependence of the moment of inertia on the angular frequency. a.} Moment of inertia $\Theta$ (in units of its classical value $\Theta_{\mathrm{cl}}$ - 
               see Methods) versus the angular frequency $\Omega$ (in units of $\Omega_{r}$), for three characteristic couplings at zero temperature. 
               At unitarity, the number of vortices entering the cloud are also reported for a few angular frequencies. 
               {\bf b.} Top view of the atomic cloud showing the array of vortices at unitarity for $\Omega/\Omega_{r}=0.3$.\end{small}}
\label{Figure-3}
\end{figure}
With the values $k_{F} \xi = 0.79$ (from ref.\cite{SPS-2013}) and $k_{F} R_{s} = 72.$ (from the present calculation) at unitarity and zero temperature, we get 
$\Omega_{c_{1}} = 0.069 \, \Omega_{r}$ where $\hbar \Omega_{r} = E_{F} / 39.9$ is the trap radial angular frequency from ref.\cite{Grimm-2011}.
This value of $\Omega_{c_{1}}$ lies within the boundaries of the full numerical calculation (see Supplementary Information),
which draws the \emph{phase diagram} for the temperature dependence of the lower critical frequency $\Omega_{c_{1}}$ about which the first vortex stably appears in the trap and of the upper critical frequency $\Omega_{c_{2}}$ about which the superfluid region disappears from the trap, in analogy to a type-II superconductor.

Figure \ref{Figure-3}a shows the moment of inertia $\Theta = L / \Omega$ obtained from the calculation of the total angular momentum $L$ (see Methods) as a function of $\Omega$ at $T=0$ for three couplings across unitarity for the geometry of ref.\cite{Grimm-2011}.
The rapid rise of $\Theta$ past the coupling-dependent threshold $\Omega_{c_{1}}$ (which decreases from the BEC to the BCS side of unitarity) is due to the progressive presence of vortices in the trap for increasing $\Omega$.
In particular, the plot reports also the number of vortices at unitarity  for a few values of $\Omega$, demonstrating that not too many vortices are needed to stabilize $\Theta$ to its classical value.
The inset of Fig. \ref{Figure-3}a amplifies the behavior of $\Theta$ at unitarity in a narrow frequency range about $\Omega_{c_{1}}$, where a smooth increase is found to occur \emph{before} the sharp rise at $\Omega_{c_{1}}$ when the first vortex nucleates.
While this effect is essentially suppressed on the BEC side, the increase of $\Theta$ for $\Omega < \Omega_{c_{1}}$ becomes more evident toward the BCS side due to the progressive presence in the outer part of the cloud of the OBP phase discussed above.
A similar effect occurs also in nuclei (where it is referred to as \emph{the yrast effect}) owing to the finite particle number \cite{Casten-1990,Bertsch-1999,Schuck-2004,Kavoulakis-2013}.
With the total particle number $N = 6 \times 10^{5}$ from ref.\cite{Grimm-2011}, this effect is expected to be of the order of $1/N \approx 10^{-5}$ which indeed corresponds to the values reported in the inset of Fig. \ref{Figure-3}a.
In addition, the linear increase of $\Theta$ before $\Omega_{c_{1}}$ obtained here is in line with a general argument provided in ref.\cite{Bertsch-1999}.

In ref.\cite{Grimm-2011}, the (nominal) angular frequency was estimated at unitarity to be $0.3 \, \Omega_{r}$, which is one order of magnitude larger than the threshold 
$\Omega_{c_{1}}$ for the nucleation of vortices obtained by our calculation.
For $\Omega = 0.3 \, \Omega_{r}$ our calculation predicts, in fact, that about $20$ vortices enter the cloud at unitarity and zero temperature, as shown in Fig. \ref{Figure-3}b.
In ref.\cite{Grimm-2011}, on the other hand, the nucleation of vortices was not considered to occur on the basis of a mechanism of resonant quadrupole mode excitation. 
This absence of vortices might be connected with a transient configuration captured by the experiment, while a longer time scale would be required to reach the situation of full thermodynamic equilibrium which is assumed by our calculation where vortices nucleate just past $\Omega_{c_{1}}$. 

\begin{figure}[t]
\includegraphics[angle=0,width=8.0cm]{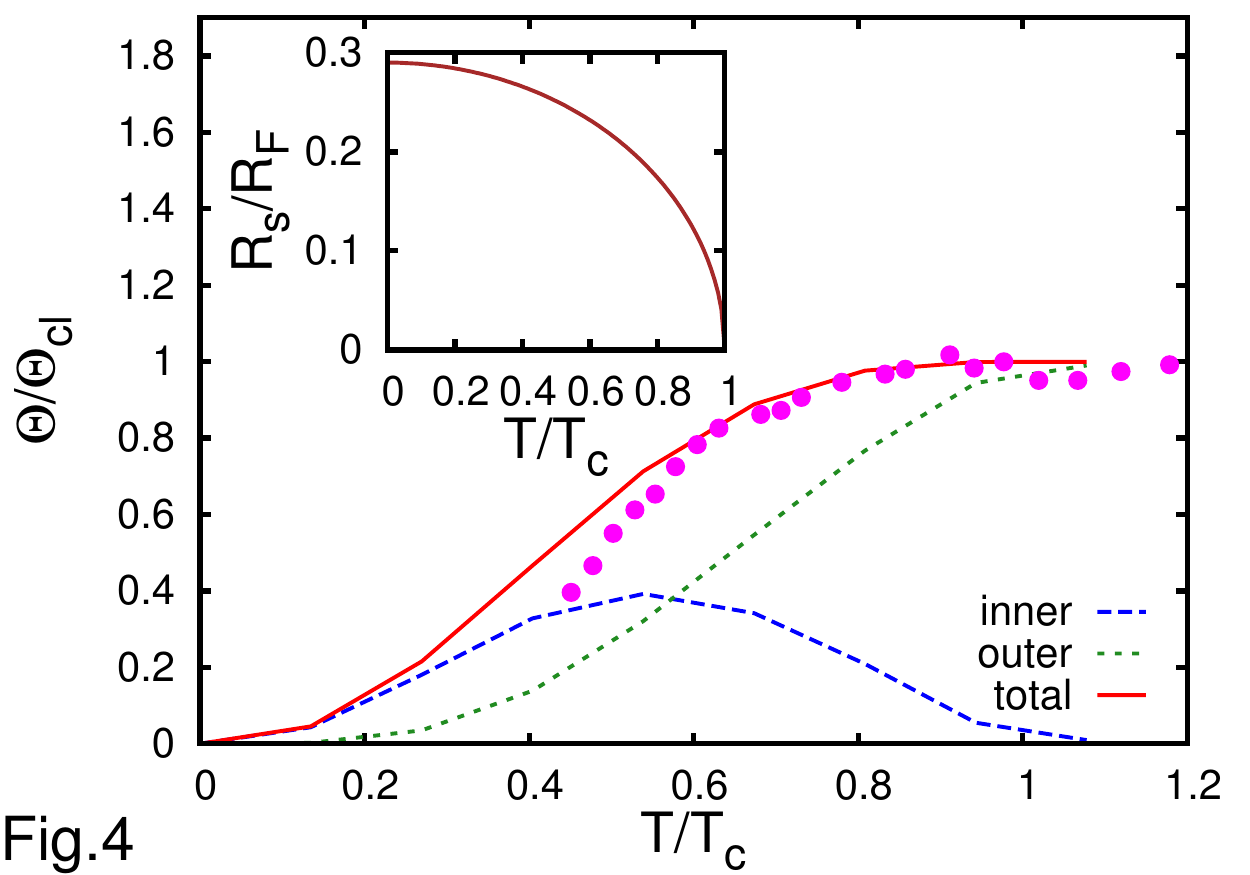}
\caption{\begin{small}{\bf Comparison with the experimental data for the moment of inertia at unitarity.} The moment of inertia $\Theta$ (in units of the classical value 
              $\Theta_{\mathrm{cl}}$) vs the temperature $T$ (in units of the critical temperature $T_{c}$) obtained from the present calculation (full line) is compared with the experimental 
               data of ref.\cite{Grimm-2011} (dots). 
               Reported are also the partial moments of inertia contributed by the inner (dashed line) and outer (dotted line) regions of the cloud. 
               The inset shows the temperature dependence of the radius $R_{\mathrm{s}}$ (in units of $R_{F}$) which sets the boundary 
               between the inner and outer regions.\end{small}}
\label{Figure-4}
\end{figure}

Accordingly, we compare the data of ref.\cite{Grimm-2011} to our calculations for the moment of inertia at unitarity for 
$\Omega = 0.003 \, \Omega_{r} \ll \Omega_{c_{1}}$, in such a way that no vortex appears in the trap at all temperatures below $T_{c}$.
This comparison is reported in Fig. \ref{Figure-4}, where $\Theta/\Theta_{\mathrm{cl}}$ is shown versus  $T/T_{c}$.
Even though our calculated value of $T_{c}$ ($=0.37 \, T_{F}$) differs from that estimated in ref.\cite{Grimm-2011} ($\simeq 0.21 \, T_{F}$),  rescaling the temperature by the respective values of $T_{c}$ leads to quite a close agreement between theory and experiment.
This kind of rescaling is a common practice in condensed matter and was recently considered also for ultra-cold Fermi gases, when comparing  theoretical predictions with  experimental data for the temperature dependence of the superfluid fraction \cite{Innsbruck-Trento-2013}.
Reported in Fig. \ref{Figure-4} are also the partial moments of inertia contributed by the superfluid (inner) and normal (outer) regions of the atomic cloud, such that $\Theta = \Theta_\mathrm{inner} + \Theta_\mathrm{outer}$.
In the inner region (with $|\mathbf{r}| \le R_{\mathrm{s}}$) the superfluid and normal components coexist with each other at a given temperature, while in the outer region 
(with $|\mathbf{r}| > R_{\mathrm{s}}$) only the normal component survives.
The corresponding temperature dependence of $R_{\mathrm{s}}$ is shown in the inset of Fig. \ref{Figure-4}.
We have verified that the moment of inertia in the inner region where $\Delta(\mathbf{r}) \ne 0$ is due only to the normal component which is present at non-zero temperature
(cf. Eq.(\ref{normal-density}) in Methods).
For $\Omega \rightarrow 0$, our approach reduces to the two-fluid model of ref.\cite{Urban-2005}, which was used recently in ref.\cite{BP-2013} to compare with the experimental data of ref.\cite{Grimm-2011} for $ \Theta$.
In ref.\cite{BP-2013}, however, a number of numerical approximations were considered to simplify the calculation, which are avoided in the present approach.

The results obtained here demonstrate how vortex arrays evolve from the BCS to the BEC regimes when a Fermi superfluid is constrained in a confined geometry, for increasing temperature and angular velocity or a combination of both.
Our theoretical approach has relied on a novel differential equation for the spatially varying gap parameter, which considerably reduces the storage space and computational time compared to the BdG equations, thereby advancing the current state-of-the-art standard for the self-consistent generation of complex arrays of quantum vortices under quite broad physical conditions.

\vspace{0.3cm}

\noindent
{\bf Methods}
\vspace{0.1cm}

\begin{small}
\noindent
{\bf The coupling parameter for the BCS-BEC crossover.} \, The BCS-BEC crossover is spanned by the coupling parameter $(k_{F} a_{F})^{-1}$, where $a_{F}$ is the scattering length of the two-fermion problem and $k_{F}$ the Fermi wave vector related to the (trap) Fermi energy $E_{F}$ by 
$k_{F}^{2}/(2m) = E_{F} = \Omega_{0} \, (3 N)^{1/3}$ with $\Omega_{0}  = (\Omega_{r}^{2} \, \Omega_{z})^{1/3}$.
Here, $m$ is the fermion mass, $N$ the total particle number, and $\Omega_{r}$ and $\Omega_{z}$ the radial and axial angular frequencies, respectively.
[Throughout the Methods, we set $\hbar$ equal unity.]
The coupling parameter ranges from $(k_{F} a_{F})^{-1} \lesssim -1$ characteristic of the weak-coupling (BCS) regime when 
$a_{F} < 0$, to $(k_{F}\, a_{F})^{-1} \gtrsim +1$ characteristic of the strong-coupling (BEC) regime when $a_{F} > 0$, 
across the value $(k_{F}\, a_{F})^{-1}=0$ at unitarity when $|a_{F}|$ diverges. 

In ref.\cite{Ketterle-2005}, the aspect ratio between the radial ($\nu_{r} = 57 \, \mathrm{Hz}$) and axial ($\nu_{z} = 23 \, \mathrm{Hz}$) trap frequencies was chosen to be about $2.5$, in order to maximize the number of produced vortices at the nominal rotation frequency $\nu = 45 \, \mathrm{Hz} \simeq 0.8 \, \nu_{r}$.
In ref.\cite{Grimm-2011}, a larger aspect ratio of about $28$ between the radial ($\nu_{r} = 680 \, \mathrm{Hz}$) and axial ($\nu_{z} = 24 \, \mathrm{Hz}$) trap frequencies was instead adopted, in order to minimize the number of produced vortices at the (nominal) rotation frequency $\nu = 200 \, \mathrm{Hz} \simeq 0.3 \, \nu_{r}$.
\vspace{0.1cm}

\noindent
{\bf LPDA equation, density, and current.} \, The LPDA equation for the gap parameter $\Delta(\mathbf{r})$ was introduced in ref.\cite{SS-LPDA}.
For the present problem of a rotating trap with angular velocity $\mathbf{\Omega}$ it reads:
\begin{equation}
- \frac{m}{4 \pi a_{F}} \Delta(\mathbf{r}) =  \mathcal{I}_{0}(\mathbf{r}) \Delta(\mathbf{r}) + 
\mathcal{I}_{1}(\mathbf{r}) \frac{\nabla^{2}}{4m} \Delta(\mathbf{r}) - i \, \mathcal{I}_{1}(\mathbf{r}) \mathbf{v}_{\mathrm{n}}(\mathbf{r}) \cdot \nabla \Delta(\mathbf{r})                                                          
\label{LPDA-differential-equation}
\end{equation}
\noindent
where
\begin{equation}
\mathcal{I}_{0}(\mathbf{r}) = \int \! \frac{d \mathbf{k}}{(2 \pi)^{3}} 
\left\{ \frac{1 - 2 f_{F}(E_{+}(\mathbf{k}|\mathbf{r}))}{2 \, E(\mathbf{k}|\mathbf{r})} - \frac{m}{\mathbf{k}^{2}} \right\}
\label{I_0-definition-finite_temperature}
\end{equation}
\noindent
and
\begin{eqnarray}
& & \mathcal{I}_{1}(\mathbf{r}) = \frac{1}{2} \int \! \frac{d \mathbf{k}}{(2 \pi)^{3}} 
\left\{  \frac{\xi(\mathbf{k}|\mathbf{r})}{2 \, E(\mathbf{k}|\mathbf{r})^{3}} \left[  1 - 2 f_{F}(E_{+}(\mathbf{k}|\mathbf{r})) \right] \right.                          
\nonumber \\
& + &  \left.  \frac{\partial f_{F}(E_{+}(\mathbf{k}|\mathbf{r}))}{\partial E_{+}(\mathbf{k}|\mathbf{r})}  \left[ \frac{\xi(\mathbf{k}|\mathbf{r})}{E(\mathbf{k}|\mathbf{r})^{2}}
- \frac{\mathbf{k}\cdot\mathbf{v}_{\mathrm{n}}(\mathbf{r})}{m \, \mathbf{v}_{\mathrm{n}}(\mathbf{r})^{2}} \frac{1}{E(\mathbf{k}|\mathbf{r})} \right] \right\} \, .
\label{I_1-definition-finite_temperature} 
\end{eqnarray}
\noindent
In these expressions, $\xi(\mathbf{k}|\mathbf{r}) = \frac{\mathbf{k}^{2}}{2m} - \mu + V(\mathbf{r})$ where $\mu$ is the chemical potential and $V(\mathbf{r})$ the external (trapping) potential, $E(\mathbf{k}|\mathbf{r}) = \sqrt{\xi(\mathbf{k}|\mathbf{r})^{2} + |\Delta(\mathbf{r})|^{2}}$,
$E_{+}(\mathbf{k}|\mathbf{r}) = E(\mathbf{k}|\mathbf{r}) - \mathbf{k} \cdot \mathbf{v}_{\mathrm{n}}(\mathbf{r})$ where 
$\mathbf{v}_{\mathrm{n}}(\mathbf{r}) = \mathbf{\Omega} \times \mathbf{r}$ is the velocity field of the normal component, and
$f_{F}(\varepsilon) = (e^{\varepsilon/k_{B}T} + 1)^{-1}$ the Fermi function. 

Correspondingly, the expression for the current density within the LPDA approach in the non-rotating frame is:
\begin{equation}
\mathbf{j}(\mathbf{r}) = \mathbf{v}_{\mathrm{s}}(\mathbf{r}) \, n(\mathbf{r})
+ 2 \int \! \frac{d\mathbf{k}}{(2 \pi)^{3}} \frac{\mathbf{k}}{m} f_{E} \! \left( E^{\mathbf{v}}_{+}(\mathbf{k}|\mathbf{r}) \right) \, .
\label{number-current-general} 
\end{equation}
\noindent
Here, $\mathbf{v}_{\mathrm{s}}(\mathbf{r}) = \nabla \varphi(\mathbf{r})/(2m)$ is the velocity field of the superfluid component and $n(\mathbf{r})$ the number density within LPDA:
\begin{equation}
n(\mathbf{r}) =  \int \! \frac{d\mathbf{k}}{(2 \pi)^{3}} \left\{ 1 - 
\frac{\xi^{\mathbf{v}}(\mathbf{k}|\mathbf{r})}{E^{\mathbf{v}}(\mathbf{k}|\mathbf{r})} 
\left[ 1 - 2 f_{F}(E^{\mathbf{v}}_{+}(\mathbf{k}|\mathbf{r})) \right] \right\} \, .
\label{number-density-general}
\end{equation}
\noindent
In the above expessions, $\xi^{\mathbf{v}}(\mathbf{k}|\mathbf{r}) = \xi(\mathbf{k}|\mathbf{r}) + \frac{1}{2} m \mathbf{v}_{\mathrm{s}}^{2} - m \mathbf{v}_{\mathrm{s}} \cdot \mathbf{v}_{\mathrm{n}}$, $E^{\mathbf{v}}(\mathbf{k}|\mathbf{r}) = \sqrt{\xi^{\mathbf{v}}(\mathbf{k}|\mathbf{r})^{2}
+ |\Delta(\mathbf{r})|^{2}}$, and $E^{\mathbf{v}}_{+}(\mathbf{k}|\mathbf{r}) = E^{\mathbf{v}}(\mathbf{k}|\mathbf{r}) 
+ \mathbf{k} \cdot ( \mathbf{v}_{\mathrm{s}}(\mathbf{r}) - \mathbf{v}_{\mathrm{n}}(\mathbf{r}) )$.

Note that the expression (\ref{number-current-general}) for the current is consistent with that of a two-fluid model (cf., e.g., ref.\cite{two-fluid-model}). 
This can be seen by expanding the argument of the Fermi function in Eq.(\ref{number-current-general}) for small values of
$|\mathbf{v}_{\mathrm{s}}(\mathbf{r}) - \mathbf{v}_{\mathrm{n}}(\mathbf{r})|$, thereby identiying the \emph{normal density} through the expression
\begin{equation}
n_{n}(\mathbf{r}) =  - \, 2 \, \int \! \frac{d\mathbf{k}}{(2 \pi)^{3}} \frac{\mathbf{k}^{2}}{3 m}
\frac{\partial f_{E}(E(\mathbf{k}|\mathbf{r}))}{\partial E(\mathbf{k}|\mathbf{r})} 
\label{normal-density}
\end{equation}
\noindent
which depends on temperature through the Fermi function.

The \emph{total angular momentum} of the system in the rotating trap is obtained in terms of the current density (\ref{number-current-general}) via the expression:
\begin{equation}
\mathbf{L} = m \int \! d\mathbf{r} \, \mathbf{r} \times \mathbf{j}(\mathbf{r}) \, .
\label{total-angular-momentum}
\end{equation}
\noindent
In the text, we have indicated by $L$ the magnitude of $\mathbf{L}$.

In addition, the moment of inertia $\Theta$ of the system is obtained from $L = \Theta \, \Omega$ and calculated from the expression 
(\ref{total-angular-momentum}) for any value of $\Omega$ (until the trapping potential can no longer sustain the atomic cloud when set into rotation).
In this respect, the present approach is not limited to small values of $\Omega$ where the formalism of linear response can be employed \cite{PS-2003}.
Under these circumstances, the moment of inertia $\Theta$ can be compared with its ``classical'' counterpart $\Theta_{\mathrm{cl}}$ defined as follows:
\begin{equation}
\Theta_{\mathrm{cl}} = m \int \! d\mathbf{r} \, n(\mathbf{r}) \, (\hat{{\Omega}} \times \mathbf{r})^{2} 
\label{classical-moment_of_inertia}
\end{equation}
\noindent
where $n(\mathbf{r})$ is the total number density given by Eq.(\ref{number-density-general}).
In particular, in the limit of vanishing angular velocity $\Theta_{\mathrm{cl}}$ coincides with the rigid-body value $\Theta_{\mathrm{rig}}$ as calculated from the density distribution, 
under the assumption that the whole cloud (including the superfluid part) performs an extremely slow rigid rotation whereby $\mathbf{v}_{\mathrm{s}} = \mathbf{v}_{\mathrm{n}} \rightarrow 0$ in Eq.(\ref{number-density-general}).

\end{small}



\begin{footnotesize}
\vspace{0.05cm}
\noindent
{\bf Acknowledgments}
\vspace{-0.02cm}

\noindent 
G.C.S. is indebted to G. Bertsch, A. Bulgac, and A. L. Fetter for discussions.
\end{footnotesize}

\newpage
\begin{center}
{\bf Supplementary Information}
\end{center}
\vspace{-0.25cm}

\begin{center}
{\bf Numerical solution of the LPDA equation}
\end{center}
\vspace{-0.25cm}

We describe the numerical procedure that we have adopted in the main text to solve the LPDA equation for the gap parameter $\Delta({\bf r})$ (cf. Eq.~(2) of the main text), together with the number equation $N=\int d{\bf r} n({\bf r})$  (with $n({\bf r})$ given by Eq.~(6) of the main text) to determine the chemical potential $\mu$ for fixed particle number $N$.  

The LPDA equation can be rewritten in the form:
\begin{equation}
\nabla^2 \Delta({\bf r}) + \frac{\tilde{{\cal I}}_0({\bf r})}{{\cal I}_1({\bf r})} \Delta({\bf r}) - 4m  i {\bf v}_n({\bf r}) \cdot \nabla \Delta({\bf r}) =0 ,
\label{LPDA2}
\end{equation}
where
$\tilde{{\cal I}}_0({\bf r})\equiv[m^2/(\pi a_F)+ {\cal I}_0({\bf r})]$ and with $\hbar = 1$. 
The coefficients ${\cal I}_0({\bf r})$ and  ${\cal I}_1({\bf r})$ are defined by Eqs.~(3) and (4) of the main text.  
The differential equation (\ref{LPDA2}) is solved over a finite box, with the condition that $\Delta({\bf r})$ vanishes identically from the boundary of the box outwards. 
For the trap of ref.~[S1] the box width is taken as 2.7$R_F$ along the $x$ and $y$ directions and 6.8$R_F$ along the $z$ direction;  
for the trap of ref.~[S2] the corresponding widths are 1.5$R_F$  and 22$R_F$. 
In this way, the box is at least 1.5 times larger than the size of the cloud in each direction for all rotation frequencies that we have considered. 

The differential equation (\ref{LPDA2}) is transformed into a set of finite-difference equations which we schematize in vector form as $\vec{F}(\vec{\Delta})=0$,  
by discretizing it over a uniform spatial grid.
Here, $\vec{\Delta}$ is a vector formed by the unknown variables $\Delta_j \equiv\Delta({\bf r}_j)$ where ${\bf r}_j$ is a point on the grid, while the equation $F_i=0$ corresponds to the finite difference version of Eq.~(\ref{LPDA2}) at position ${\bf r}_i$.
For the trap of ref.~[S1] the grid is taken  $800 \times 800 \times 30$ for the $x,y$, and $z$ directions, respectively; while for the trap of ref.~[S2] the grid is taken  $500 \times 500 \times 50$.
One is thus left with solving a system of $N_p\simeq 2 \times 10^7$ non-linear equations for the $N_p$ complex variables $\Delta_j$, where the non-linearity arises from the functional dependence of the coefficients ${\cal I}_0({\bf r})$ and ${\cal I}_1({\bf r})$ on $\Delta({\bf r})$ (cf.~Eqs.~(3) and (4) of the main text). 
To solve this system we have implemented a quasi-Netwon method as follows. 
The ordinary (multi-dimensional) Newton method would imply modifying $\vec{\Delta}$ as follows:
\begin{equation}
\vec{\Delta}^{\rm new}=\vec{\Delta}^{\rm old} - {\bf J}^{-1} \cdot \vec{F}(\vec{\Delta}^{\rm old})
\label{update}
\end{equation}
where ${\bf J}^{-1}$ is the inverse of the Jacobian matrix ${\bf J}$ with matrix elements $J_{ij}=\frac{\partial F_i(\vec{\Delta}^{\rm old})}{\partial \Delta_j}$. 
Here, the Jacobian matrix can be calculated quite accurately with a numerical effort comparable to that of evaluating $\vec{F}(\vec{\Delta})$, because most of the off-diagonal elements vanish and those different from zero are linear combinations of $\Delta_{i}$ and $\Delta_j$.  
The memory storage of the sparse matrix ${\bf J}$ is set up by using a compressed sparse column (CSC) format. 
However, since the numerical inversion of ${\bf J}$ is too costly, we have resorted to an incomplete $LU$ factorization~[S3] and obtained an approximate inverse of ${\bf J}$ to be 
inserted in Eq.(\ref{update}). 
It is the use of this approximate inverse of ${\bf J}$ that makes the procedure a quasi-Newton method instead of an ordinary Newton method.  
Specifically for our problem, this method proves to converge better than alternative versions of the quasi-Newton method (such as the SR1 or Broyden's methods~[S4]).

For a given trial value of $\mu$, we routinely perform 40 iterations for the discretized gap $\vec{\Delta}$ according to Eq.~(\ref{update}). 
We then update the chemical potential $\mu$ through a single step of the secant method applied to the number equation $N=\int d{\bf r} \, n({\bf r})$ at fixed $\Delta({\bf r})$. 
With this new value of $\mu$, we again repeat 40 iterations for $\vec{\Delta}$, and so on.  
In the presence of a large number of vortices (about one hundred or more), 50 steps to update $\mu$, each followed by 40 iterations for $\vec{\Delta}$, are typically required to reach a satisfactory convergence. 
With a smaller number of vortices, on the other hand, these numbers can considerably be decreased (together with the number of points for the spatial grid in the $xy$ plane).    
In order to speed up the calculation (and to make it feasible, in practice, for a large number of vortices), the coefficients $\tilde{{\cal I}}_0({\bf r})$ and ${\cal I}_1({\bf r})$ are calculated over an interpolation grid $100\times100\times100$ in the variables ($|{\bf A}|$, $|\Delta|$, $\bar{\mu}\equiv \mu - V({\bf r})$), with a logarithmic spacing for $\Delta$. 
[Note that this grid is over the possible values of $|{\bf A}|$, $|\Delta|$, and $\bar{\mu}$, and not over the physical space spanned by the variable ${\bf r}$.]  
The values of  the coefficients $\tilde{{\cal I}}_0({\bf r})$ and ${\cal I}_1({\bf r})$ at position ${\bf r}$ are then obtained by a trilinear interpolation within a cube containing the point $|{\bf A}({\bf r})|$, $|\Delta({\bf r})|$, and $\bar{\mu}({\bf r)}$.  
In this way, the most demanding cases (like that shown in Fig. 1a of the main text) required 30 hours of CPU time on a standard desktop computer (with no parallelization of the code). 

For frequencies close to $\Omega_{c_1}$, where the solution with one vortex is almost degenerate with that without vortices, one needs to be particularly careful. 
In this case, we have used for the initial {\em ansatz} the product of the gap profile $\Delta_{TF}(\mu - V(x,y,z))$ within a local density approximation for the system in the absence of  rotation ~[S5] (where $V(x,y,z)$ is the trapping potential), with the function 
\begin{equation}
f({\bf r};{\bf R}_{0})= \frac{x-X_{0} + i(y- Y_{0}) }{\sqrt{\xi_{0}^{2} + (x- X_{0})^{2} + (y - Y_{0})^{2}}}
\label{f_r}
\end{equation}
which simulates a vortex of radius $\xi_{0}$ centered at ${\bf R}_{0}=(X_{0},Y_{0})$. 
Here, the position and radius of the vortex are parameters that can be varied to optimize the solution. 
In particular, the initial position of the vortex should be slightly displaced from the trap center, to avoid the system being trapped in an excited state for $\Omega < \Omega_{c_1}$. 
In this way, for $\Omega > \Omega_{c_1}$ (but still close to $\Omega_{c_1}$) the vortex adjusts its position and radius to the convergence values.
For $\Omega <  \Omega_{c_1}$, on the other hand, the vortex migrates towards the edges of the cloud and eventually disappears.

At larger rotation frequencies, to speed up the calculations we have used as initial \emph{ansatz} a gap profile given by the above local density approximation, but now multiplied by a triangular lattice of vortices which are spaced according to Feynman's theorem. 
Specifically, this is implemented by multiplying the gap profile $\Delta_{TF}(\mu - V(x,y,z))$ within local density with the function 
$\prod_{\mathrm{v}=1}^{\mathcal{N}_{\mathrm{v}}} \, f({\bf r};{\bf R}_{\mathrm{v}})$, where $f({\bf r};{\bf R}_{\mathrm{v}})$ is given by the expression (\ref{f_r}) with ${\bf R}_{\mathrm{v}}=(X_{\mathrm{v}},Y_{\mathrm{v}})$ replacing ${\bf R}_{0}=(X_{0},Y_{0})$ and $\xi_{\mathrm{v}}$ replacing $\xi_{0}$, $\{ (X_{\mathrm{v}},Y_{\mathrm{v}}) ; \mathrm{v} = 1, \cdots, \mathcal{N}_{\mathrm{v}}  \}$ being the initial positions of 
$\mathcal{N}_{\mathrm{v}}$ vortices in a triangular lattice spaced according to Feynman's theorem.
It turns out that this initial configuration contains about twice the number of vortices of the final configuration at convergence (since Feynman's theorem is progressively violated when approaching the border of the cloud).  
Correspondingly, the iteration procedure evolves in such a way that a number of vortices progressively evaporates away from the border of the cloud, until the system reaches its final equilibrium configuration. 
Quite generally, in the course of the iterations the radii of the vortex cores adjust to their convergence values rather quickly, while a larger number of iterations is required to reach convergence as far as the positions and number of vortices are concerned.  

\begin{center}
{\bf Spontaneous self-assembling of vortex arrays}
\end{center}
\vspace{-0.25cm}

The distinct power of the present method is that vortex arrays can be generated through the cycles of self-consistency at a finite rotation frequency, even when the initial condition for the gap profile contains (essentially) no vortices. 
In this respect, during iterations we have found that, quite generally, vortices enter from the edges of the cloud and eventually reach their equilibrium positions inside the cloud. 

As a demonstration of a typical numerical simulation that shows the evolution through the cycles of self-consistency, we report here a simplified version of the calculations presented in the main text, where now it is the chemical potential $\mu$ and not the particle number $N$ to be kept fixed at a given value through the cycles of self-consistency (in this case, we use thermodynamic value $\mu = 0.752 E_{F}$ of Fig. 3b of the main text).
In addition, to speed up the calculation further we now utilize a spatial grid with $300 \times 300 \times 35$ points instead of the denser one 
with $500 \times 500 \times 50$ points used for the calculations in the main text for the trap of ref.~[S2].

\begin{figure}[ht!]
\centering
\includegraphics[width=80mm]{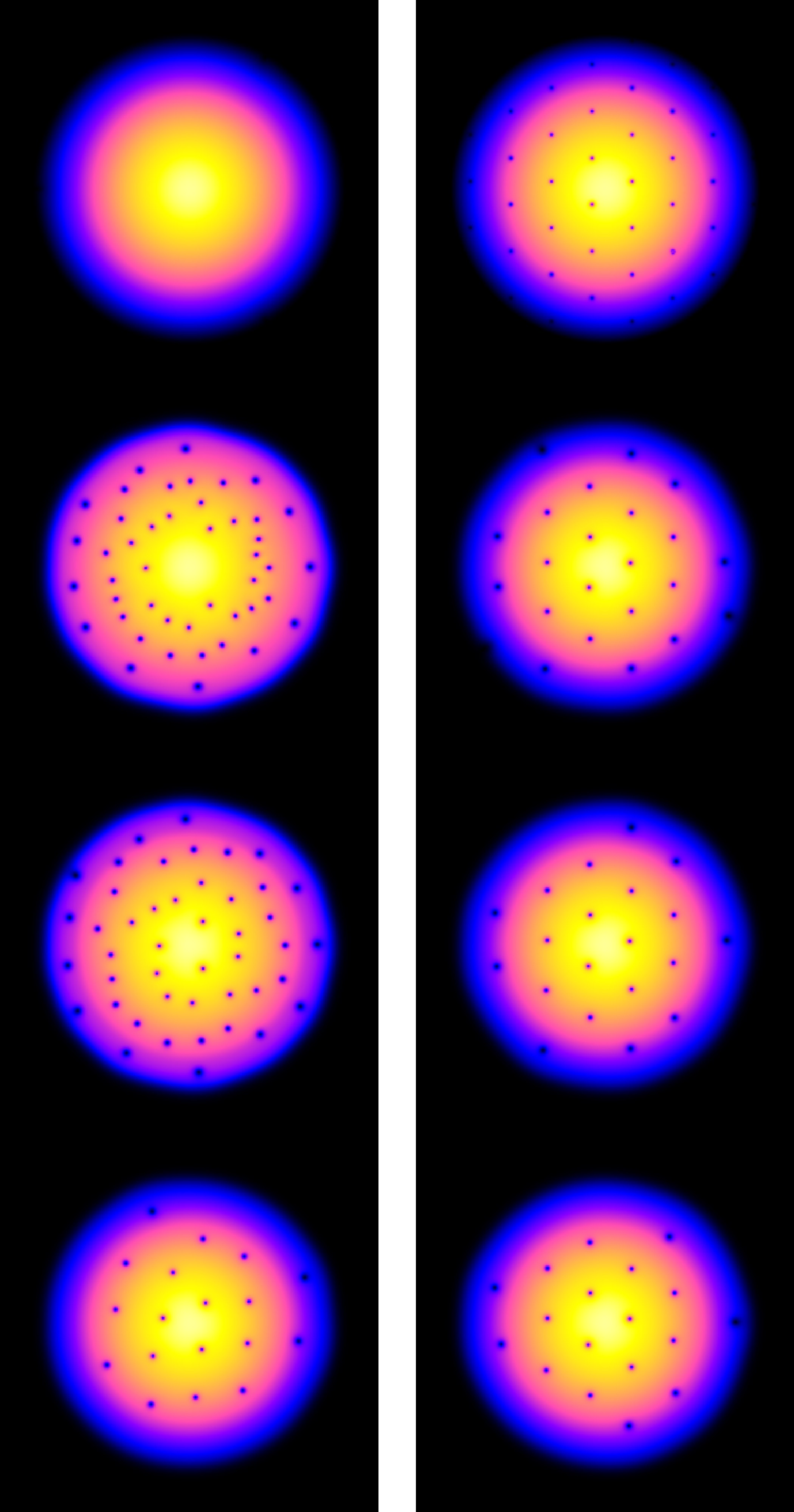}
\caption{{\bf  $\;$ The evolution of the gap profile through iterations.} The evolution of the gap profile in the course of the iterations is shown at unitarity, $T=0$, 
              and $\Omega=0.3 \Omega_r$ for the trap parameters of ref.~[S2]. 
              Left column (from top to bottom): initial configuration without vortices, and configurations after 400, 800, and 18000 iterations. 
              Right column (from top to bottom): initial configuration with 37 vortices, and configurations after 400, 800, and 6000 iterations.}
\end{figure}

Figure S1 shows a typical evolution of the gap profile during the cycles of self-consistency for the trap of ref.~[S2] at unitarity, zero temperature, and $\Omega=0.3 \Omega_r$, using \emph{two different initial configurations}:  
in the left column, a gap profile with no vortex but only a phase imprint of equilateral triangular symmetry at the cloud edge; 
in the right column, a gap profile containing 37 vortices as required by Feynman's theorem. 
In order to reduce the distortion of the triangular lattice in the left column, the equilibrium value $0.3 \Omega_r$ has been reached only asymptotically in the course of the iteration cycles through a suitable damped saw-tooth profile of the angular frequency $\Omega$
(a movie showing the complete evolution for this case is available at 
\colorbox{cyan}{\color{blue}{\href{http://bcsbec.df.unicam.it/?q=node/1}{http://bcsbec.df.unicam.it/?q=node/1}}\color{black}}).
The result is that these two quite different initial configurations lead essentially (apart from an overall rotation) to the same final solution at convergence with a total of 18 vortices 
(note also that the panel at the bottom of the right column of Fig. S1 coincides with Fig. 3b of the main text, the minor differences being ascribed to the different number of points in the spatial grids).

\begin{center}
{\bf The $\Omega$ vs $T$ phase diagram for the superfluid phase of a neutral trapped Fermi gas}
\end{center}
\vspace{-0.25cm}

It is interesting to combine together the two physical effects which yield a finite value for the moment of inertia in the superfluid phase, namely, the presence of: 
(i) An array of vortices even in the absence of a normal component when the angular frequency increases above a threshold, as occurs at zero temperature (cf. Fig. 3a of the main text);  
(ii) A normal component at finite temperature even in the absence of vortices, as occurs for vanishing angular frequency 
(cf. Fig. 4 of the main text).
Simultaneous consideration of both effects leads us to construct a \emph{phase diagram} for the temperature dependence of the lower critical frequency $\Omega_{c_{1}}$ about which the first vortex stably appears in the trap
and of the upper critical frequency $\Omega_{c_{2}}$ about which the superfluid region disappears from the trap.
This phase diagram is the analogue of that for a homogeneous type-II superconductor, showing the temperature dependence of the critical magnetic fields $H_{c_{1}}$ and 
$H_{c_{2}}$ ~[S6]. 

\begin{figure}[t]
\begin{center}
\hspace{-0.65cm}
\includegraphics[angle=0,width=8.0cm]{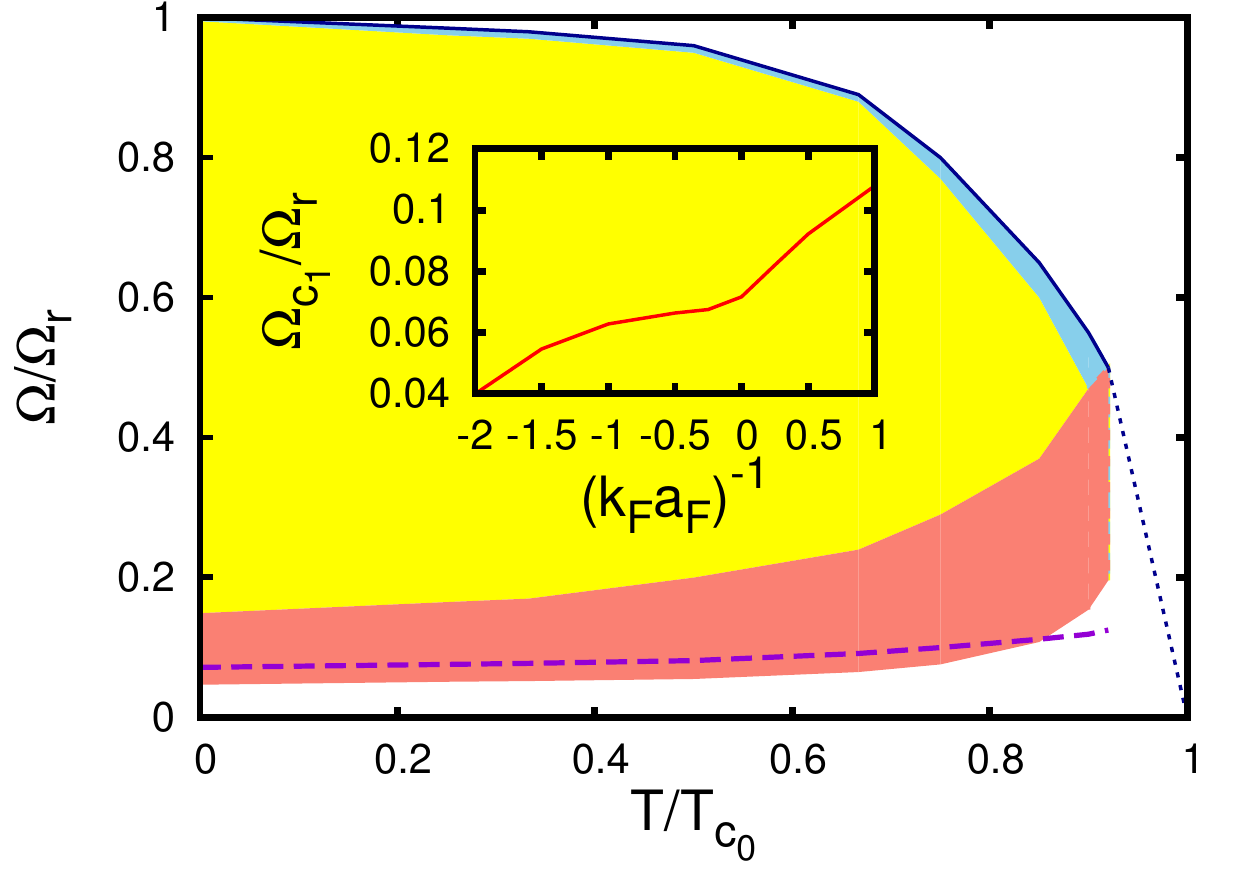}
\caption{{\bf $\;$ Phase diagram for the critical frequencies $\Omega_{c_{1}}$ and $\Omega_{c_{2}}$ vs temperature.} 
               Temperature dependence of the two critical frequencies $\Omega_{c_{1}}$ and $\Omega_{c_{2}}$ (in units of the radial trap frequency $\Omega_{\mathrm{r}}$) 
               at unitarity for the trap of ref.~[S2]. 
              The shaded red (blue) area corresponds to the uncertainty associated with $\Omega_{c_{1}}$ ($\Omega_{c_{2}}$).              
              At a given temperature, no vortex is present for $0 \le \Omega < \Omega_{c_{1}}^{(l)}$ while vortex arrays appear for 
              $\Omega_{c_{1}}^{(u)} \le \Omega < \Omega_{c_{2}}^{(l)}$ (shaded yellow area). 
              The broken line extrapolates the values of $\Omega_{c_{2}}^{(u)}$ down to the point $(\Omega=0,T=T_{c_{0}})$, while the dashed line corresponds to Eq.(1) 
              of the main text with appropriate (temperature dependent) values of $\xi$ and $R_{s}$.
              The inset shows $\Omega_{c_{1}}$ at $T=0$ vs the coupling $(k_{F} a_{F})^{-1}$ again obtained from Eq.(1) of the main text with the appropriate values of $\xi$ 
              and $R_{s}$ at zero temperature.}
\end{center}
\end{figure}

The results of this calculation are reported in Fig. S2 at unitarity for the trap corresponding to the experiment of ref.~[S2]. 
More precisely, at a given temperature we have found it necessary to distinguish between a lower ($l$) and an upper ($u$) value for 
$\Omega_{c_{1}}$ and for $\Omega_{c_{2}}$ according to the following considerations.
The lower value $\Omega_{c_{1}}^{(l)}$ corresponds to the smaller angular frequency $\Omega$ at which an isolated vortex placed initially close to the trap center (say, at a distance $R_{\mathrm{s}}/10$ from it) begins to be attracted toward the trap center in the course of the cycles of the self-consistent solution of the LPDA equation.
The upper value $\Omega_{c_{1}}^{(u)}$ corresponds instead to the smaller value of $\Omega$ at which an isolated vortex placed initially at the edge $R_{\mathrm{s}}$ of the superfluid part of the cloud begins to be attracted toward the trap center.
The ensuing uncertainty in the identification of $\Omega_{c_{1}}$, which is unavoidably present for a trap with finite size, corresponds to the shaded red area of Fig. S2.
On the other hand, the lower value $\Omega_{c_{2}}^{(l)}$ corresponds to the smaller value of $\Omega$ at which all vortices have eventually disappeared from the trap, while the upper value $\Omega_{c_{2}}^{(u)}$ is determined by the condition that the gap parameter itself vanishes everywhere in the trap.
[In this context, we have found that approaching $\Omega_{c_{2}}^{(u)}$, the spatial width of $\Delta(\mathbf{r})$ shrinks progressively but never becomes smaller than the size of the ground-state wave function of the harmonic trap, and that from this point on it is the height of $\Delta(\mathbf{r})$ to decrease to zero.]
The ensuing uncertainty in the identification of $\Omega_{c_{2}}$ corresponds to the shaded blue area of Fig. S2.
For the specific trap and coupling conditions under which Fig. S2 was constructed, the values of $\Omega_{c_{1}}$ and $\Omega_{c_{2}}$ with their related uncertainties could be identified up to the maximum temperature $0.92 \, T_{c_{0}}$ where $T_{c_{0}}$ is the critical temperature in the trap for $\Omega=0$.
Finally, the shaded yellow area which extends from $\Omega_{c_{1}}^{(u)}$ to $\Omega_{c_{2}}^{(l)}$ is where arrays of vortices are present for given values of $\Omega$ and $T$.

For comparison, the dashed line in Fig. S2 corresponds to the approximate expression (1) of the main text, in which we have inserted the appropriate (temperature-dependent) values of $k_{F} \xi$ for an isolated vortex taken from ref.~[S7] and of $k_{F} R_{s}$ obtained from the present calculation.
It is remarkable that the curve obtained from the approximate expression (1) of the main text falls within the shaded red area of Fig. S2 obtained by the full numerical calculation.
We have then used this approximate expression to get an estimate for the coupling dependence of $\Omega_{c_{1}}$ at zero temperature which is reported in the inset of Fig. S2.
The values of $\Omega_{c_{1}}$ obtained in this way are in line with the sharp thresholds occurring in Fig. 3a of the main text, where, however, the numerical procedure sets these thresholds at the lower values of $\Omega_{c_{1}}^{(l)}$.

\begin{figure}[t]
\includegraphics[angle=0,width=8.0cm]{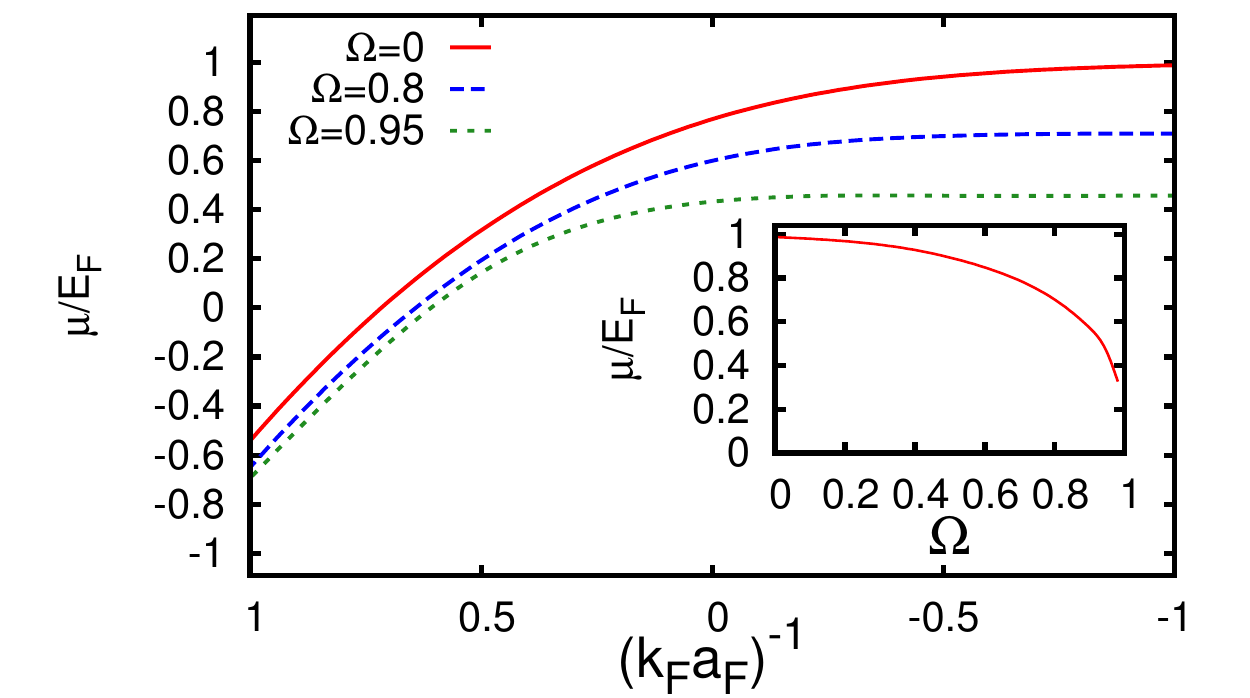}
\caption{{\bf Effect of rotation on the chemical potential.} 
               Coupling dependence of the chemical potential in the trap (in units of $E_{F}$) at $T=0$ for three different angular frequencies (in units of  $\Omega_{r})$. 
               The inset shows the chemical potential vs $\Omega$ for $(k_{F} a_{F})^{-1} = -1$.}
\label{Figure-S3}
\end{figure}

\begin{center}
{\bf Chemical potential}
\end{center}
\vspace{-0.15cm}

Figure S3 shows the coupling dependence of the chemical potential $\mu$ in the trap at zero temperature for several angular frequencies approaching the limiting value $\Omega = \Omega_{r}$, past which the fermion cloud is no longer bound.
The rotation affects $\mu$ more markedly on the BCS than on the BEC side of unitarity, the dependence becoming rather abrupt in the BCS limit as shown in the inset of 
Fig. S3 for $(k_{F} a_{F})^{-1} = -1$. 

\begin{figure}[h]
\begin{center}
\hspace{-0.65cm}
\includegraphics[angle=0,width=8.0cm]{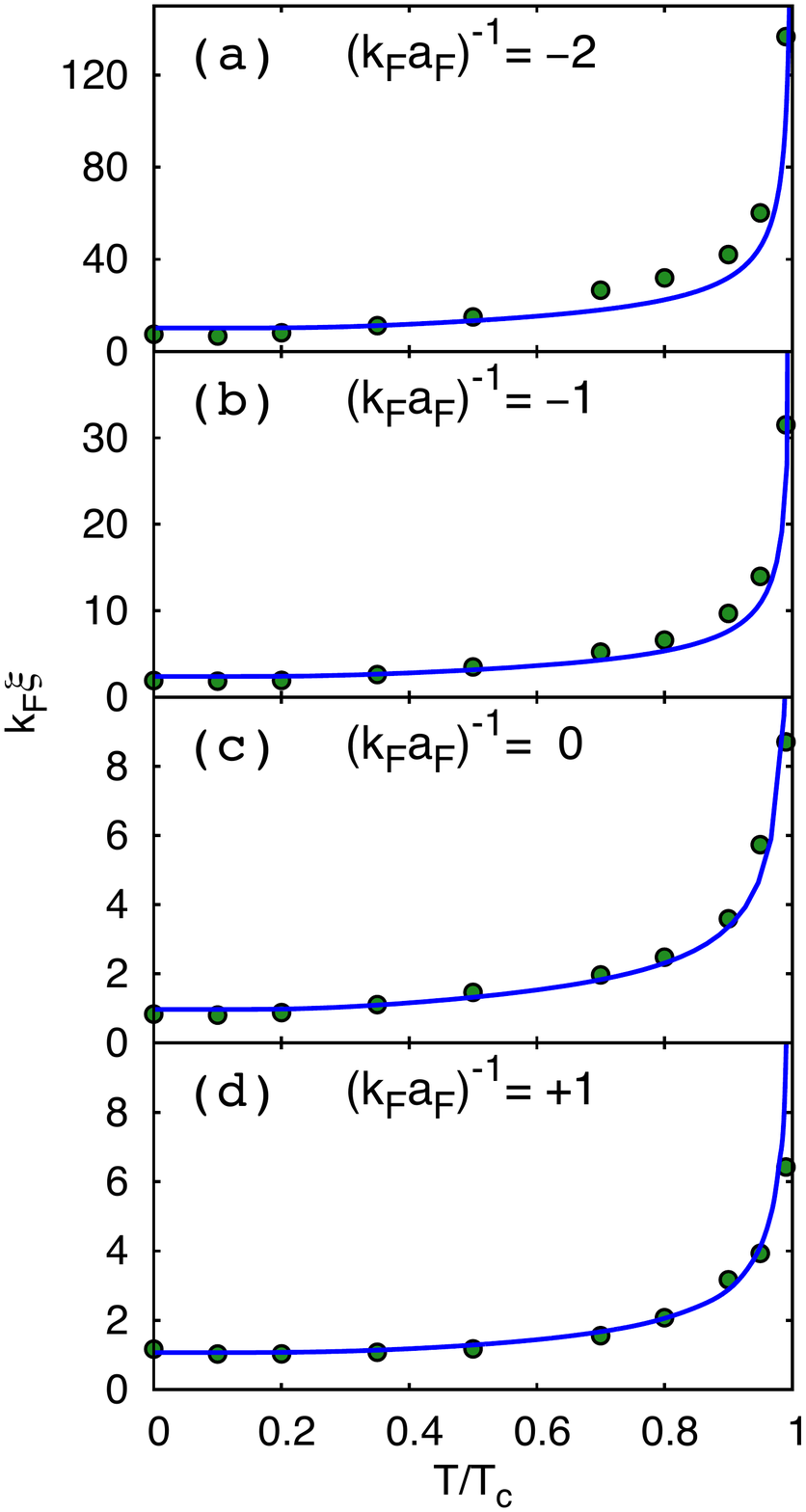}
\caption{{\bf $\;$ Coherence length obtained by homogeneous mean field plus Gaussian fluctuations and by inhomogeneous mean field.} 
Comparison of the temperature dependence of the coherence length $\xi$ (in units of $k_{F}^{-1}$), obtained alternatively by including Gaussian fluctuations on top of the homogeneous BCS mean field (full lines) and by a numerical solution of the inhomogeneous BdG equations for an isolated vortex embedded in an infinite superfluid (dots), for different values of the coupling parameter $(k_{F} a_{F})^{-1}$.
The temperature is in units of the superfluid critical temperature $T_{c}$ of the homogeneous system.
The results of the homogeneous calculation have been rescaled by an overall factor of $2/3$, which takes into account the different definitions used for the same physical quantity by the two independent numerical calculations. [Figure adapted from Fig. 10 of ref.~[S9].]}
\end{center}
\end{figure}

\begin{center}
{\bf Fluctuation corrections emerging from an inhomogeneous mean-field approach}
\end{center}
\vspace{-0.15cm}

\begin{figure}[hb]
\begin{center}
\hspace{-0.65cm}
\includegraphics[angle=0,width=8.0cm]{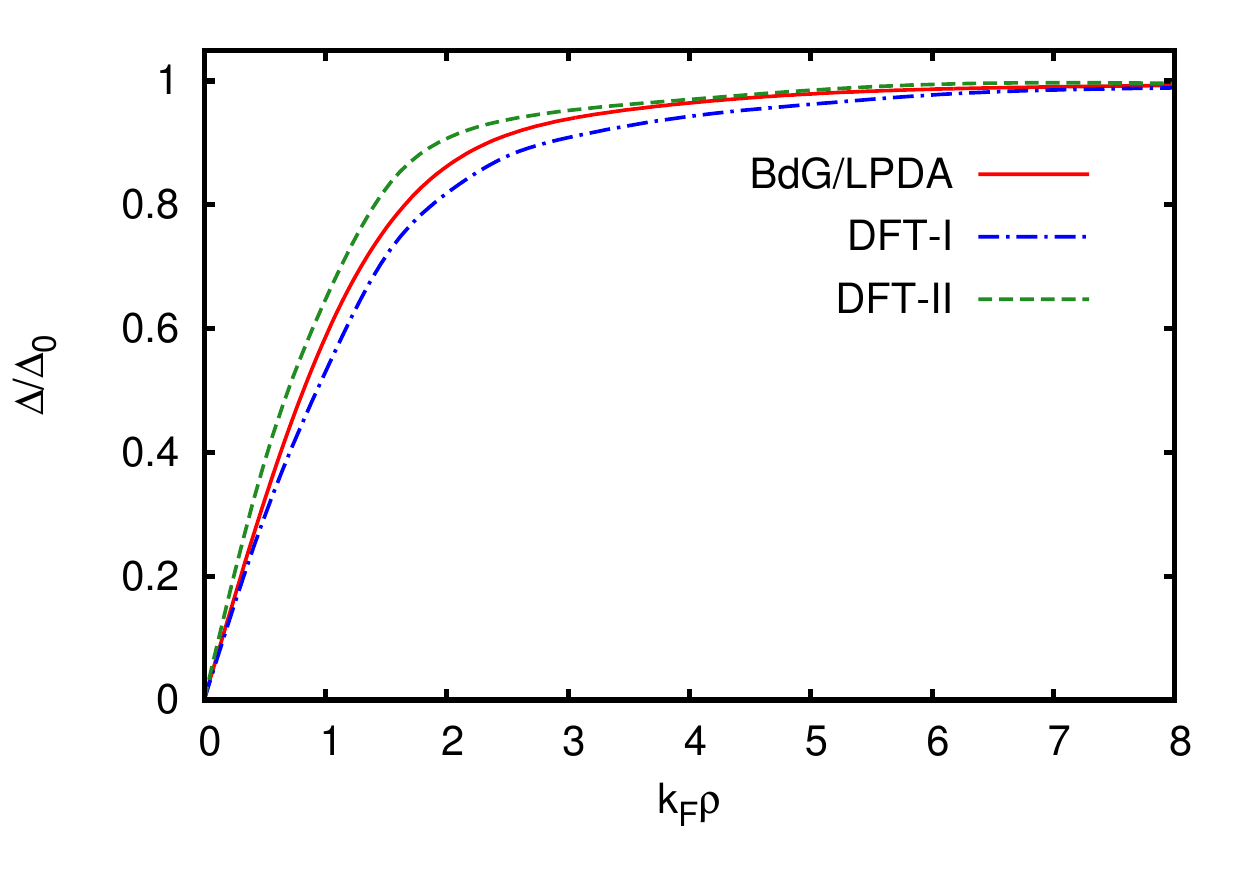}
\caption{{\bf $\;$ Effect of fluctuation corrections on the profile of an isolated vortex.} Comparison of the vortex profiles at unitarity and zero temperature, obtained by the BdG/LPDA approach (full line) and by the DFT approach (dashed-dotted and broken lines).
The BdG/LPDA profile is taken from Fig.~1c of the main text, while the DFT profiles have been extracted from Fig.~2 of ref.~[S10] where two different parameterizations (I and II) were used for the DFT approach (referred to as the EDF approach in that reference).}
\end{center}
\end{figure}

Quite generally, a mean-field calculation for an inhomogeneous situation (of the type dealt with by the BdG or LPDA equations) contains contributions from what are referred to as fluctuation corrections in a homogeneous situation.
This is because, in an inhomogeneous situation, the imprint of the lowest excited states can be found in the ground-state wave function (as discussed, for instance, in ref.~[S8]).
As an example, Fig.~S4 reports a comparison of the coherence (healing) length of the gap parameter, obtained alternatively by solving the BdG equations for an isolated vortex 
(cf. ref.~[S7]) and by adding pairing (Gaussian) fluctuations on top of the homogeneous BCS mean field (cf. ref.~[S9]).
This comparison shows that a BdG calculation is able to capture fluctuation contributions beyond mean field as far as the spatial variations of the gap parameter are concerned.

In addition, as emphasized in the main text, the vortex profile (and thus the healing length) is not affected by the presence of the surrounding vortices, and the distance between two adjacent vortices is an order of magnitude larger than the healing length. 
Possible corrections to the healing length should thus have a minimal impact on the distribution of vortices.

Nor even the further inclusion of pairing fluctuations beyond the Gaussian ones is expected to change the vortex profile significantly. 
This is shown in Fig.~S5, which compares the vortex profiles at unitarity and zero temperature, obtained alternatively by the BdG/LPDA approach (full line) and by the Density-Functional-Theory (DFT) approach of ref.~[S10] (dashed-dotted and broken lines) within two different parameterizations (referred to as I and II).
Rather remarkably, the BdG/LPDA profile just lies within the uncertainty of the DFT profiles in this important case ~[S11].

\vspace{0.3cm}
\noindent
{\bf References}
\vspace{0.1cm}

\begin{small}

\noindent
[S1] M. W. Zwierlein, J. R. Abo-Shaeer,  A. Schirotzek, C. H. Schunck, and W. Ketterle, Nature {\bf 435}, 1047 (2005).\\ 
\noindent
[S2] S. Riedl, E. R. S\'anchez Guajardo, C. Kohstall, J. Hecker-Dencshlag, and R. Grimm, New J. Phys. {\bf 13}, 035003 (2011).\\
\noindent
[S3]  C. G. Broyden and M. T. Vespucci, {\em Krylov Solvers for Linear Algebraic Systems} (Elsevier B.V., Amsterdam, 2004). \\ 
\noindent
[S4]  J. Nocedal and S. J. Wright, {\em Numerical Optimization} (Springer Science, New York, 1999).\\
\noindent
[S5] A. Perali, P. Pieri, and G. C. Strinati, Phys. Rev. A {\bf 68}, 031601 (2003).\\
\noindent
[S6] P. G. de Gennes, \emph{Superconductivity of Metals and Alloys} (Benjamin, New York, 1966). \\
\noindent
[S7] S. Simonucci, P. Pieri, and G. C. Strinati, Phys. Rev. B {\bf 87}, 214507 (2013).\\
\noindent
[S8]  E. P. Gross, J. Math. Phys. {\bf 4}, 195 (1963).\\
\noindent
[S9] F. Palestini and G. C. Strinati, Phys. Rev. B {\bf 89}, 224508 (2014).\\
\noindent
[S10] A. Bulgac and Y. Yu, Phys. Rev. Lett. {\bf 91}, 190404 (2003).\\
\noindent
[S11] At unitarity, the profile of an isolated vortex obtained by the LPDA approach coincides at any temperature with that obtained by the BdG approach, as
                          shown in Fig. 2 of ref.~[S12].\\
\noindent
[S12] S. Simonucci and G. C. Strinati, Phys. Rev. B {\bf 89}, 054511 (2014).  

\end{small}

\end{document}